\documentclass{article}
\usepackage{spconf,amsmath,graphicx}
\usepackage{hyperref}
\title{Privacy Sensitive Speech Analysis \\ using Federated Learning to Assess depression}
\name{Suhas BN
and Saeed Abdullah%\sthanks{saeed@psu.edu}
}
\address{College of Information Sciences \& Technology\\ Pennsylvania State University, University Park, USA\\
{$\left\{\text{suhas,saeed}\right\}\text{\hspace{-0.1cm}@psu.edu}$}}
\begin{document}
%\ninept
%
\maketitle
\begin{abstract}
Recent studies have used speech signals to assess depression. However, speech features can lead to serious privacy concerns. To address these concerns, prior work has used privacy-preserving speech features. However, using a subset of features can lead to information loss and, consequently, non-optimal model performance. Furthermore, prior work relies on a centralized approach to support continuous model updates, posing privacy risks. This paper proposes to use Federated Learning (FL) to enable decentralized, privacy-preserving speech analysis to assess depression. Using an existing dataset (DAIC-WOZ), we show that FL models enable a robust assessment of depression with only 4--6\% accuracy loss compared to a centralized approach. These models also outperform prior work using the same dataset. Furthermore, the FL models have short inference latency and small memory footprints while being energy-efficient. These models, thus, can be deployed on mobile devices for real-time, continuous, and privacy-preserving depression assessment at scale.

\end{abstract}
\section{Introduction}
\vspace{-0.2cm}
Depression is a severe mental illness that affects millions of people worldwide \cite{Depression_WHO}. Depression causes staggering individual and societal costs, including a higher risk of mortality \cite{Depression_WHO}. There remains a significant treatment gap — individuals with depression often do not receive adequate treatment \cite{thornicroftUndertreatment2017}.

Addressing this treatment gap requires effective detection and monitoring of depression at scale. Toward this goal, recent studies have established that speech features including prosodic, articulatory, and acoustic signals can be indicative of depression onset and severity \cite{cumminsreview2015}. However, speech signals can be highly privacy-sensitive, and continuous monitoring can be challenging. Specifically, privacy-sensitive analysis is critical for assessing mental health issues. As such, recent studies have focused on using only privacy-preserving speech features (e.g., \cite{wyatt2011inferring,abdullah2016automatic}). However, using a subset of speech features leads to considerable information loss, negatively impacting model accuracy. Also, prior work requires a centralized approach to support continuous model updates from collected data. The need for data sharing across individuals, even for a subset of features, can lead to serious privacy risks. There has been an increasing focus on enabling decentralized training to reduce privacy risks in recent years. For example, Federated Learning (FL) aims to train models using local datasets and then merge those local models as necessary. Given that local datasets are not shared, FL significantly reduces privacy concerns. However, the resultant model accuracy and overhead can be a concern, particularly when deploying in mobile devices. To the best of our knowledge, there has been no work to establish the accuracy and performance overhead of using FL for privacy-preserving speech analysis to assess depression. 

This paper aims to address this gap in two steps. First, we use an existing dataset (DAIC-WOZ) to establish that decentralized, privacy-preserving models can be used for robust assessment of depression using speech data. Second, we establish that the computational overhead of these models is low, and thus, they can be deployed to mobile devices for continuous and real-time monitoring. The paper makes the following novel technical contributions:
\vspace{-0.1cm}
\begin{itemize}
    \item We used transfer learning to assess depression and compared  training overhead and accuracy using two FL frameworks: Federated Averaging (FedAvg) \cite{mcmahan2017communication} and Federated Matching Averaging (FedMA) \cite{wang2020federated}.
    \vspace{-0.2cm}
    \item The FL models achieve significantly better accuracy compared to the best-performing models in prior work using the DAIC-WOZ dataset (e.g., 87\% accuracy for the combined dataset compared to 74.64\% in \cite{srimadhur2020end}).
    \vspace{-0.2cm}
    \item We deploy the models in a smartphone to assess performance overhead for determining depression and show real-time and continuous assessment is possible.
\end{itemize}

\vspace{-0.3cm}
\section{Related Work}
\vspace{-0.3cm}
%\subsection{Speech and Depression}
Prior work has found that depressive states correspond to changes in prosodic and acoustic features (e.g., reduced pitch range, loudness, energy dynamics, and speaking rate) \cite{ cumminsreview2015}. There has been an increasing focus on using speech as an objective biomarker of depression. Long et al. \cite{long2017detecting} used different speech features (e.g., short-time energy, intensity, formant frequencies, shimmer, jitter, and ZCR) on an in-house dataset and achieved 78.02\% in detecting depression. Lalitha et al. \cite{lalitha2019enhanced} used deep neural networks to identify speech features that contain ``emotional information". They achieved an average accuracy of 84.3\% on the Berlin EmoDB database.

%\subsubsection{DAIC-WOZ dataset}
While DAIC-WOZ includes audio and video streams, we will focus on audio data, given the paper's scope. In recent work, Ma et al. \cite{ma2016depaudionet} developed DepAudioNet, which leveraged Deep Convolutional Neural Networks and Long Short-Term Memory networks. The model achieved an F1 score of 0.52, precision of 0.35, recall of 1, and an approximated accuracy of 0.5. Hanai et al. \cite{al2018detecting} developed an LSTM model using audio features with an accuracy of 0.59. Srimadhur et al. \cite{srimadhur2020end} used an end-to-end network for classification from audio data with an accuracy of  74.64\%. However, these previous studies do not focus on privacy-preserving audio analysis.

%\subsection{Privacy preserving speech analysis}
Prior work has explored the use of privacy-sensitive features from speech  \cite{wyatt2011inferring}. For example, Wyatt et al. \cite{wyatt2011inferring} aimed to limit intelligibility and speech content reconstruction using a subset of features. However, reducing the feature space can lead to considerable information loss and negatively impact model accuracy. Furthermore, this approach still requires a central data repository to support model training and updating. Given that depression can be a lifelong condition, it is often necessary to continuously update models over depression to reflect different personal and lifestyle changes over time. The need for uploading and sharing data, even for a subset of features, can lead to serious privacy risks.

%\subsection{Federated Learning}
\begin{figure}%[htb]
    \centering
    \includegraphics[width=\linewidth,trim={0 0 0 350},clip]{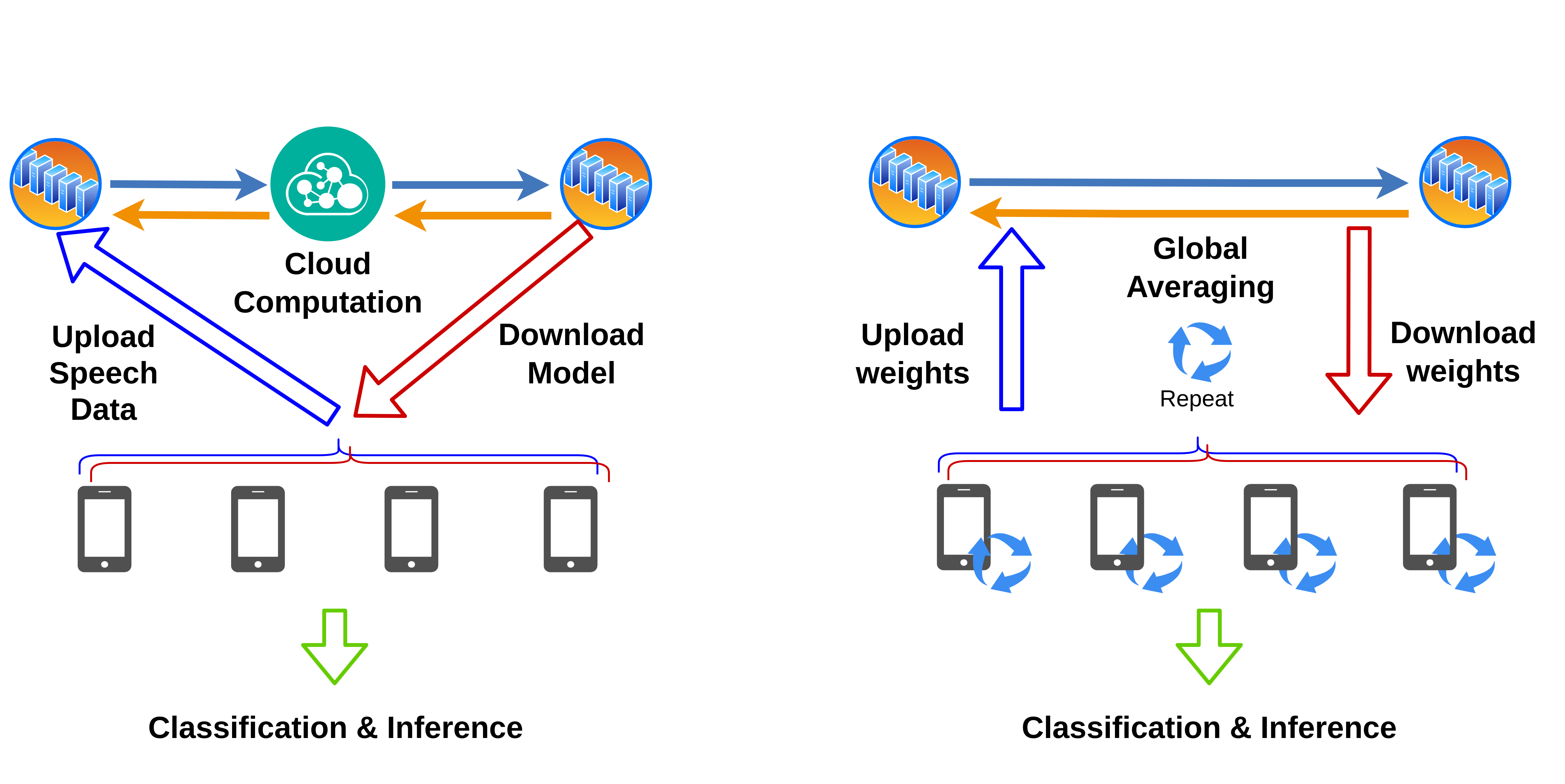}
    \vspace{-0.8cm}
    \caption{The centralized (left) and federated learning methods (right) differ in model training and updating steps. 
    %In the centralized method (left) case, the data is aggregated and uploaded to a centralized server. The features are either extracted at the source or the server. Following the feature extraction process, a model is trained in the server. The smartphones then receive the trained model, which performs classification and inference tasks. In the case of federated learning methodologies (right), the training happens on the device. Devices initially receive a global model. Each device then updates the global model based on its local dataset. The updated model weights (parameters) are then sent back to the server. These updated weights are then merged to obtain a new global model then deployed back to devices. In other words, no data leaves the device, which can significantly reduce privacy risks.
    }
    \label{overall_system_design}
    \vspace{-0.5cm}
\end{figure}

Several recent studies have used FL to reduce privacy risks in analyzing health data \cite{langlotz2019roadmap,brisimi2018federated}. Figure \ref{overall_system_design} compares the centralized and federated learning approaches. FL requires a trade-off regarding model performance and training overhead. Centralized models have access to more data, leading to better performance. The on-device training and merging model parameters can lead to considerable training time in FL. These trade-offs can be critical for systems focusing on real-time and continuous assessment of mental health issues, including depression. While FL has been used for different health applications, there is a lack of existing work that leveraged FL for privacy-preserving speech analysis to assess depression. This paper aims to address this gap. Specifically, we aim to establish a benchmark of accuracy and performance overhead of using FL to assess depression using speech signals from the DAIC-WOZ dataset. 

\begin{table}[]
    \centering
    \begin{tabular}{ccc}
    $\begin{array}{ccc}
\hline \text { Gender } & \text { hasDepression} & \text { noDepression } \\
\hline \text { Male } & 25 & 25 \\
\text { Female } & 25 & 25 \\
\hline
\text { Total } & 50 & 50 \\
\hline
\end{array}$
    \end{tabular}
    \vspace{-0.3cm}
    % \caption{The dataset used in this paper. It has been ensured that the dataset is balanced across experiments.}
    \caption{The dataset used in this paper. We randomly selected a subset of the DAIC-WOZ dataset to ensure equal gender and depression class distribution.}
    \label{tab:dataset}
\end{table}

\vspace{-0.3cm}
\section{Data Attributes and Preprocessing}
\begin{figure}%[htb]
    \centering
    \includegraphics[width=\linewidth,trim={0 0 0 0},clip]{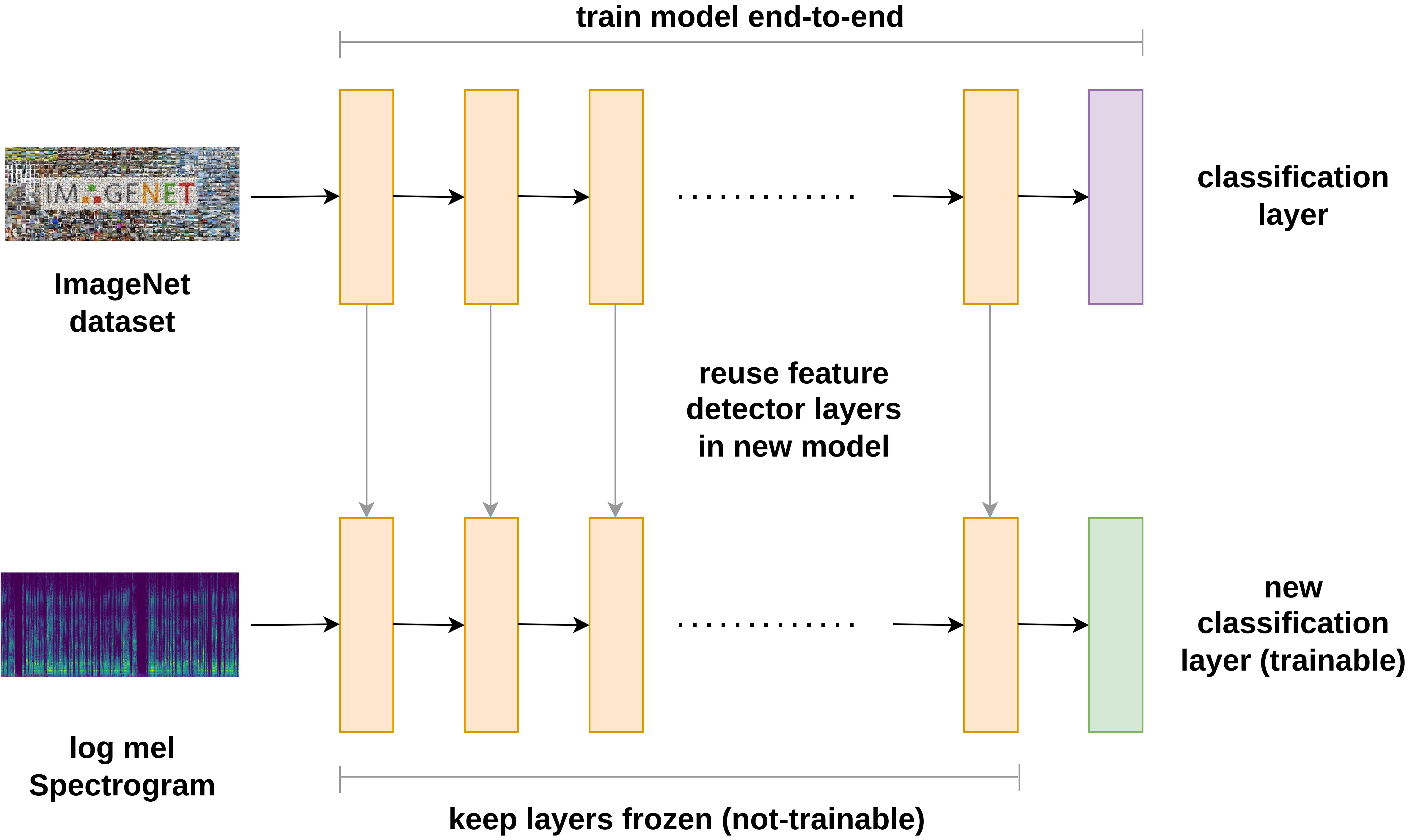}
    \vspace{-0.4cm}
    \caption{We used transfer learning strategies to efficiently train models. We started with pre-trained models and reused their feature detector layers. We then trained the output layer using log mel spectrograms from the dataset.}
    
    % \caption{The basic system design of this work uses the transfer learning methodology wherein we reuse the feature detector layers of the three pre-trained models while training the output layer using log mel spectrograms. 
    %The three models (ResNet-18, GoogleNet, and MobileNet v2) are trained on the ImageNet dataset (top). ImageNet dataset contains 50,000 different classes that help networks trained on them understand patterns quickly. The next step involves reusing the feature detector layers of the already trained model (pre-trained network) while keeping the output layer trainable. By doing this, the network quickly learns new design patterns from the log mel spectrogram images. We also reduce the training required to classify log mel spectrogram features  (bottom).
    %}
    \vspace{-0.5cm}
    \label{TransferLearning}
\end{figure}

\vspace{-0.4cm}
\label{Dataset_Preprocessing}
We have used the DAIC-WOZ dataset \cite{gratch2014distress} to train and evaluate model performance. The dataset contains 189 interviews. Each interview has a corresponding PHQ-8 \cite{PHQ8} score indicating depression severity. The original dataset is imbalanced regarding gender and the number of individuals with depression. To ensure balance, we randomly selected a subset of the dataset with equal gender and depression class distribution (25 individuals each) as shown in Table \ref{tab:dataset}. The speech data has a sampling rate of 16 kHz. It contains speeches from both the participant and the virtual interviewer. The dataset provides start and stop timestamps indicating where the interviewer and the participant have spoken. We use this metadata to extract speech segments for the participants. Following prior work \cite{suhas2020speech}, we used an overlapping window length of 1s duration with a shift of 0.1s to extract log spectrogram features using the scipy.signal.spectrogram function, which is log-scaled as shown in Fig.  \ref{TransferLearning}. The images help model both temporal and harmonic structures of audio signals, leading to improved classification performance over existing methods.

% \begin{figure}[ht] 
%   \label{Ch3-figure:t-SNE} 
%     \centering
%     \includegraphics[width=\linewidth]{images/tsne_2class.png} 
%     \caption{t-SNE plot of the data considered in this work with different distance measures}
%     \label{fig:2class_tSNE}
%     \vspace{4ex}
% \end{figure}

%%%%%%%%%%TRIAL

%%%%%%%%%%%%%%%%%%%%%%%%%%%%%%%%%%%%%%%%%%%%%%

\begin{table}[]
    \centering
    \begin{tabular}{cc}
    $\begin{array}{cccc}
\hline \text { Network } & \text { Depth } & \text { Parameters (M) } & \text { Input} \\
\hline 
\text{GoogleNet} & 22 & 7 & 224 \times 224 \\
\text{MobileNet v2} & 20 & 3.4 & 224 \times 224 \\
\text{ResNet-18} & 18 & 11.7 & 224 \times 224 \\
\hline
\end{array}$
    \end{tabular}
    \vspace{-0.4cm}
    \caption{The characteristics of the network architectures used}
    \vspace{-0.5cm}
    \label{tab:architectures}
\end{table}

\section{Experimental Setup}
\vspace{-0.3cm}
\label{Ch:4.2_Experimental_Setup}
We used PyTorch on an NVIDIA Tesla V100-SXM2-32GB
GPU to run the experiments. We use the Stochastic Gradient Descent optimizer (optim.SGD) with an initial learning rate (LR) = 0.001 and momentum of 0.9 for training the three networks. The `LR' value is decayed every 7 epochs by a factor of gamma = 0.1 \cite{you2019does}. This decaying of the LR is useful in the case of FedAvg as it ensures convergence guarantees \cite{li2019convergence}. There has been work \cite{jeong2018communication,yao2018two} to alleviate high communication costs and performance drop, especially in the case of \textbf{non-I.I.D scenario} \cite{li2020federated,nagaraj2021privacy}. Also, instead of training a single model on devices in FL environments, the authors suggest using a two-stream model, widely used in transfer learning  \cite{long2015learning}. We employ the Cross entropy function (nn.CrossEntropyLoss) as loss criterion. We minimized the risk of overfitting by using the five-fold cross-validation and early stopping criteria. The spectrogram features extracted are of dimensions 515x389 and are corrected to 224x224 using the transforms library from torchvision - where only the training images are augmented. Given our focus on using these models in smartphone devices, we chose three pre-trained models with a smaller number of parameters  ($<$ 12 Million) as in Table \ref{tab:architectures}.

%%%%%%%%%%%%%%%%%%%%

\vspace{-0.25cm}
\section{Findings}
\vspace{-0.3cm}
  In this work, we focus on two binary classification tasks, namely - Depression classification (i.e., hasDepression vs. noDepression) and Depression severity classification (i.e., mild vs. severe). We consider a centralized and two FL approaches (FedAvg and FedMA). For FL approaches, we consider five devices on the same network.

\vspace{-0.25cm}
\subsection{Depression classification}
\begin{figure}
\centering
\centerline{\includegraphics[width=\linewidth,trim={80 30 245 0},clip]{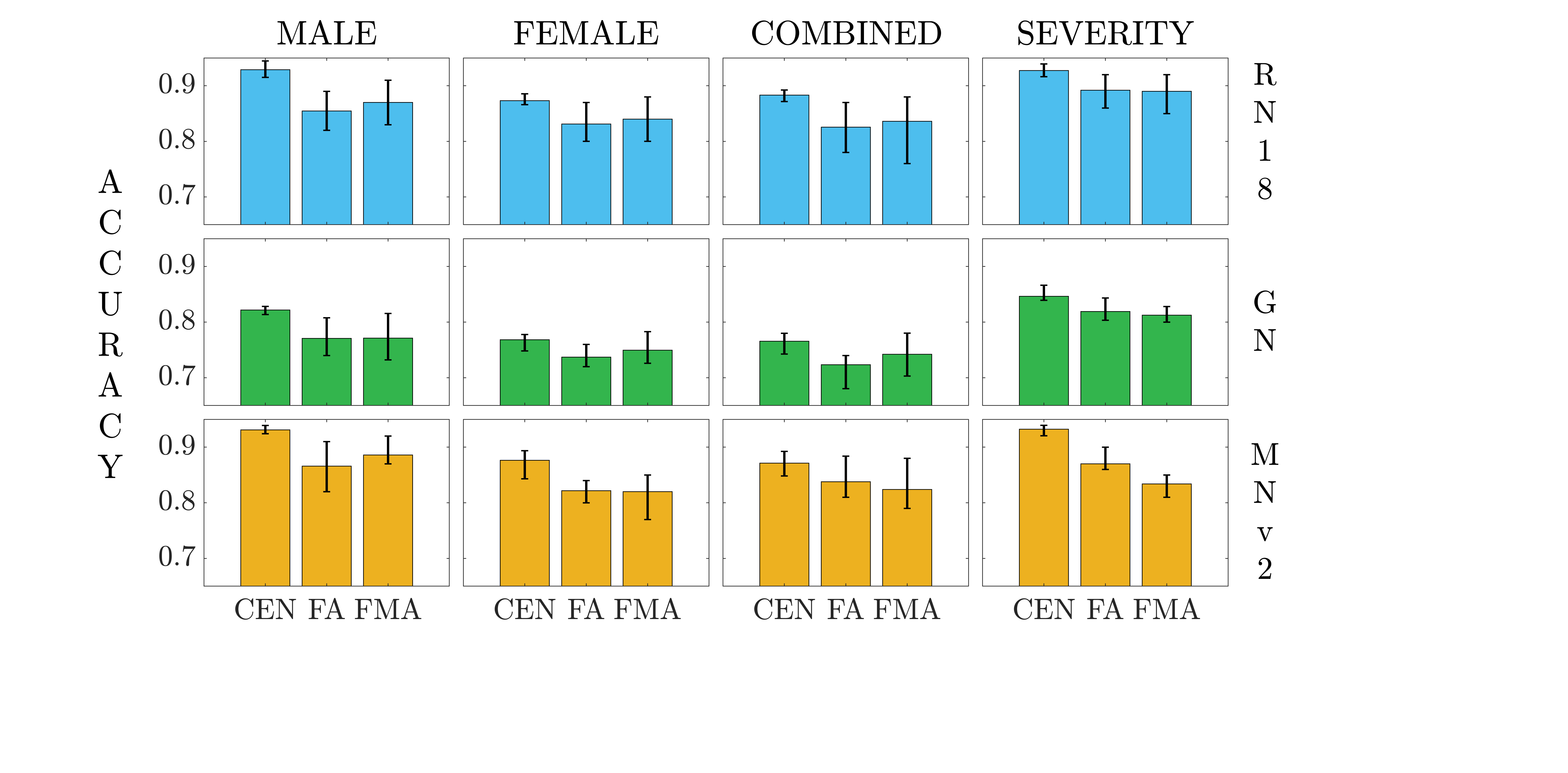}}
\vspace{-1.2cm}
%\centerline{{(a) Male}}
\label{fig:test1}
%\caption{Average five-fold cross-validation accuracy (has vs. no depression) for different algorithms (bars), networks (rows), and subjects (columns).}
\caption{Average five-fold cross-validation accuracy (has vs. no depression) for different algorithms (CEN: centralized, FA: FedAvg, FMA: FedMA). We also calculated performance across networks (rows) and subjects (columns).}
\vspace{-0.5cm}
%It is consistently seen across graphs that the centralized algorithm performs best compared to the two federated algorithms. We observe both FedAvg and FedMA performing equally well overall, albeit in different experiments among the federated algorithms. This can be seen in Table \ref{tab:Ch5:results-1}. We minimized the risk of overfitting by using the five-fold cross-validation and early stopping criteria. Furthermore, we have used the evaluation guideline established in prior work \cite{saebVoodoo2016a} to prevent overestimation of model accuracy. 
\label{fig:binary-depression-classification}
\end{figure}

\vspace{-0.2cm}
For the depression classification task, we first assess model performance for gender-specific data. We then consider the combined dataset for model performance for comparison. We use five-fold cross-validation throughout the classification tasks. The resultant model performance is shown in Figure \ref{fig:binary-depression-classification}. A summary of the average training overhead is in Table \ref{tab:all-time}. 
\begin{table}[]
    \centering
    \begin{tabular}{ccccc}
    $\begin{array}{ccccc}
\hline \text {Type} & \text {Male} &\text {Female}&\text {Combined}&\text {Severity}\\
\hline \text{RN-18} & 155&167&253&222\\
\text{GN} & 235&255&411&216\\
\text{MN v2} & 204&223&359&191\\
\hline \text{RN-18 + FA} & 340 &372&574&499\\
\text{GN + FA} & 380&457&633&475\\
\text{MN v2 + FA} & 345 &415&595&411\\
\hline \text{RN-18 + FMA} & 327&487&554&470\\
\text{GN + FMA} & \text{} 366&480&612&466\\
\text{MN v2 + FMA} & \text{} 344&399&570&419\\
\hline
\end{array}$
    \end{tabular}
\vspace{-0.4cm}
    \caption{Average time taken (s) to train different networks and algorithms. Note: RN-18 (ResNet-18), GN (GoogleNet), MNv2 (MobileNet v2), FA (FedAvg), FMA (FedMA)}
    \label{tab:all-time}
    \vspace{-0.2cm}
\end{table}
Figure \ref{fig:binary-depression-classification} compares model performance for all the four sets of experiments. The centralized approach performs better than the federated methods by 6-10\% across folds. The best avg. five-fold accuracy for the centralized approach is 0.934, while for the federated scheme, it is 0.91. From Table \ref{tab:all-time}, we see that the centralized method is about 1.55-2.19x faster than federated schemes with ResNet-18 fastest for both centralized (155s) and federated schemes (327 \& 340s respectively). Regarding the female subjects, the best average five-fold accuracy for centralized is 0.89, while for the FL scheme, it is 0.87. The centralized approach is about 1.78-2.91x faster than federated schemes. The centralized approach performs better by 4-6\% across folds for the combined dataset. The best five-fold accuracy for the centralized approach is 0.885, while for FL, it is 0.87. The centralized method is about 1.48-2.26x faster than federated schemes.

%%%%%%%%%%%%%%%%%%%%%%%%%%%%%%%%%%%%%%%%%%%%%%%%%%%

\vspace{-0.25cm}
\subsection{Depression severity classification}
\vspace{-0.2cm}
Classifying mild and severe states can lead to informed clinical decision-making and pre-emptive care. We used the PHQ-8 score $>$10 as the threshold. This resulted in a binary classification task with low (PHQ-8 score $<$10) and high (PHQ-8 $\geq$10). The low class includes noDepression as well (i.e., PHQ-8 $\leq$4). We used the combined dataset for the classification task, resulting in a reasonably balanced class distribution (low: 28 subjects and high: 22 subjects). The results are shown in Figure \ref{fig:binary-depression-classification}. The best average five-fold accuracy in centralized and FL schemes are 0.93 and 0.88 (both in ResNet-18), respectively. The centralized method is about 2.11-2.24x faster than federated schemes, with MobileNet v2 training the fastest for both centralized and FL schemes (see Table \ref{tab:all-time}).

\vspace{-0.25cm}
\subsection{Overall performance comparison}
\vspace{-0.25cm}
We used a ranking method to summarize overall performance across different approaches and networks. For each task category (e.g., Male), we assign a rank (1–3) to rate performance with rank 1 for the best performing model. By combining ranks across tasks (12 in total), we compare the robustness and consistency of different approaches. The results are summarized in Table \ref{tab:Ch5:results-1} and \ref{tab:Ch5-results-2} respectively. The centralized algorithm takes the least time to train and has the best accuracy. FedMA (2.33) performs slightly better than FedAvg (2.67) in time, but their accuracy ranking is the same, indicating similar performance across tasks. We compare network architecture performance in Table \ref{tab:Ch5-results-2}. MobileNet v2 performs well in both time and accuracy and offers the best of both worlds.

To summarize, the privacy-preserving FL models perform robustly with only 4-6\% accuracy lost compared to a traditional centralized approach  (\emph{hasDepression} vs. \emph{noDepression}). The FL models achieve comparable accuracy to the centralized approach for assessing depression severity (\emph{high} vs. \emph{low}). These findings establish the feasibility of using privacy-preserving FL models for depression assessment.

\begin{table}[htb]
    \centering
    \begin{tabular}{lcc|lcc}
    $\begin{array}{ccc|ccc}
\hline  & \text {Time} &&&\text {Accuracy}&\\
\hline \text{Method} & \text{Sum} &\text{Avg.} &\text{Method} &\text{Sum} &\text{Avg.}\\
\hline \text{Central}&12&1&\text{Central}&12&1\\
\text{FedAvg}&32&2.67&\text{FedAvg}&30&2.5\\
\text{FedMA}&28&2.33&\text{FedMA}&30&2.5\\
\hline
\end{array}$
    \end{tabular}
    \vspace{-0.3cm}
    \caption{Performance ranking (lower is better) for different approaches across twelve different task categories.}
    \label{tab:Ch5:results-1}
    \vspace{-0.3cm}
\end{table}

\begin{table}[htb]
    \centering
    \begin{tabular}{ccc|ccc}
    $\begin{array}{ccc|ccc}
\hline  & \text {Time} &&&\text {Accuracy}&\\
\hline \text{Network} & \text{Sum} &\text{Avg.} &\text{Network} &\text{Sum} &\text{Avg.}\\
\hline \textbf{RN-18}&\textbf{6}&\textbf{1.5}&\text{RN-18}&7&1.75\\
\text{GN}&11&2.75&\text{GN}&12&3\\
\text{MN v2}&7&1.75&\textbf{MN v2}&\textbf{5}&\textbf{1.25}\\
\hline
\end{array}$
    \end{tabular}
    \vspace{-0.3cm}
    \caption{ Performance ranking for different network architectures (lower is better). The ranking is across twelve different task categories. MobileNet v2 provides the best trade-off.}
    \label{tab:Ch5-results-2}
    \vspace{-0.3cm}
\end{table}

\section{Deployment feasibility}
\vspace{-0.3cm}
\begin{figure}[htb] 
   \label{fig8} 
   \begin{minipage}[b]{0.5\linewidth}
     \centering
     \centerline{\includegraphics[width=\linewidth,trim={0 1400 0 250},clip]{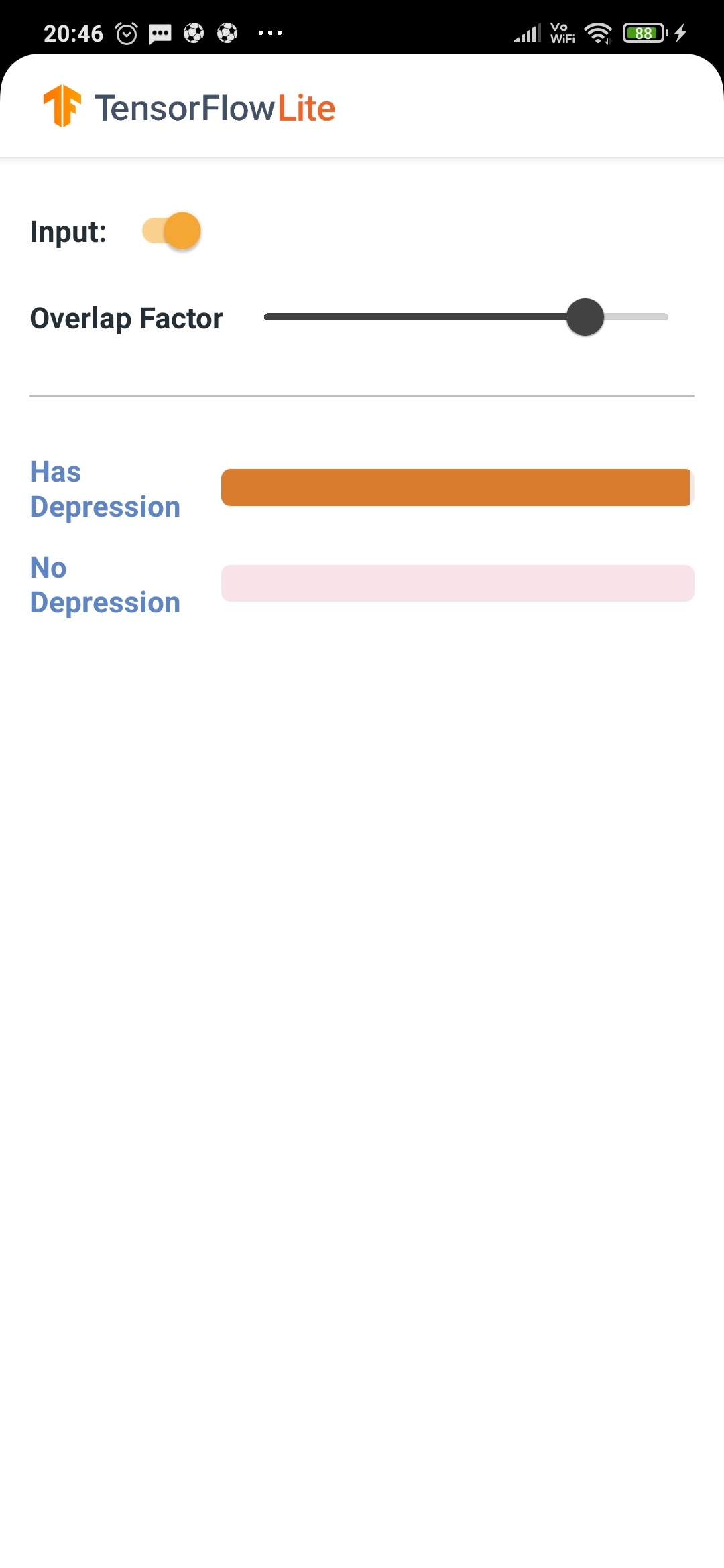}} 
     \centerline{{(a) Speech with depression}}
     \label{fig:Ch5-has}
     \vspace{4ex}
   \end{minipage}%%
   \begin{minipage}[b]{0.5\linewidth}
   \centering
     \centerline{\includegraphics[width=\linewidth,trim={0 1400 0 250},clip]{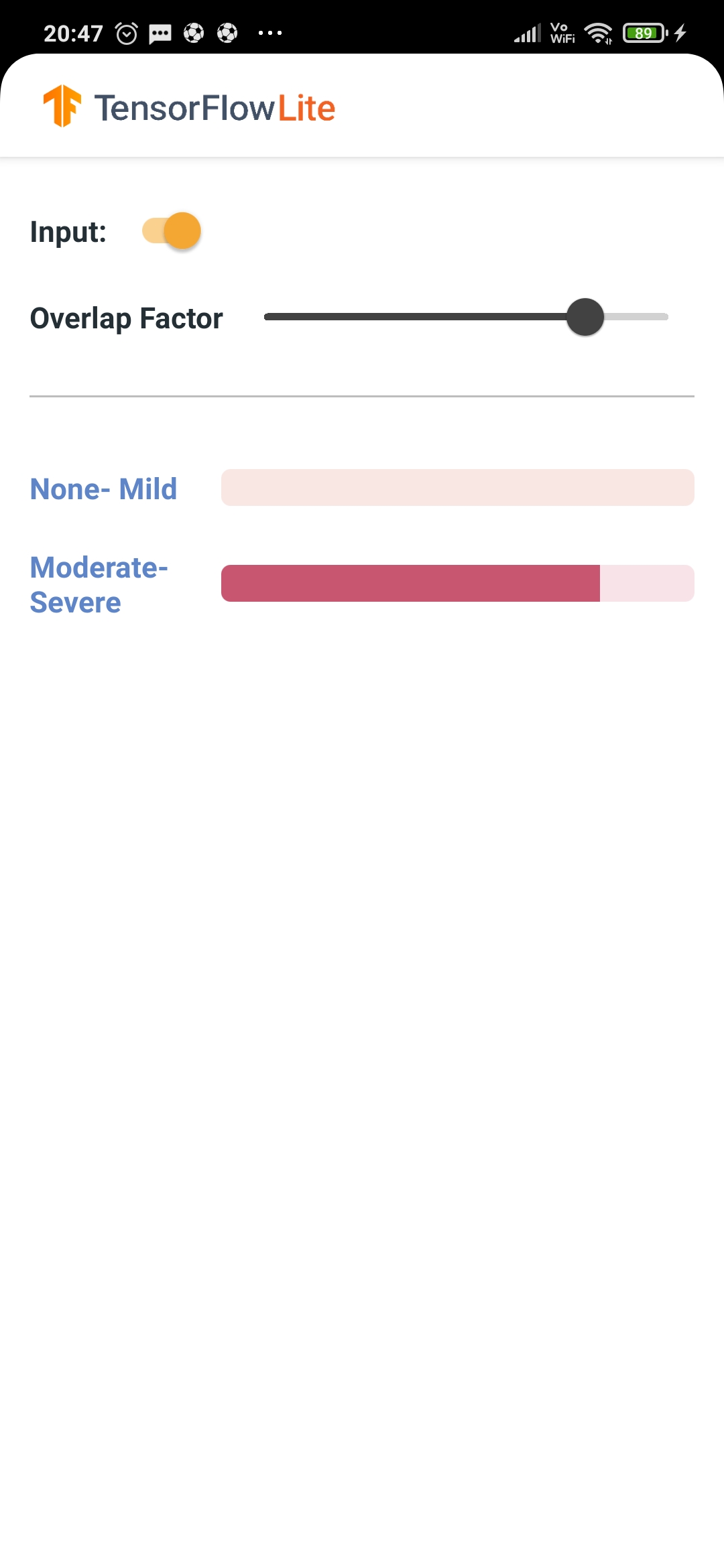}} 
     \centerline{{(b) Severity: Moderate-Severe}}
     \label{fig:Ch5-Moderate}
     \vspace{4ex}
   \end{minipage} 
  \vspace{-1.3cm}
   \caption{The developed Android app. (a) It first aims to classify whether the speech shows signs of depression, and if it does, (b) it proceeds to rate its severity.}
   \vspace{-0.3cm}
 \end{figure}
 
While the performance of the privacy-preserving, decentralized models is highly encouraging, it is critical to make sure that they can be deployed to devices with low computational resources. To assess the feasibility, we developed a smartphone app using TensorFlow Lite. The app classifies whether the speech shows signs of depression and, if it does, rates its severity. We used a Xiaomi Redmi Note 7 with 4GB RAM/64 GB ROM running Android 11 for deployment. For classification, we used an optimal two-fold model approach using the combined model and a model from the subject's gender (in real-time). Since the data capture phase is common and training takes little time - thus ensuring that the app works in real-time. Once we confirm depression exists through majority voting of the individual spectrogram image frames for both models, we use a depression severity model from the same model group. Thus, classification is more robust compared to a one-size-fits-all model. We used Android Profiler to collect real-time CPU, memory, network, and battery consumption data. The findings are shown in Table \ref{tab:app-all-time}. For all models, the inference latency $<$ 100ms, which is essential for in-situ and real-time assessment. Each model requires $<$ 50MB of memory. The models are energy efficient --- energy consumed/frame ranges from 0.26-0.37 joules, which translates to 72-102 µW/hr.

\begin{table}[htb]
    \centering
    \begin{tabular}{cccc}
    $\begin{array}{cccc}
\hline \text {Type} & \text {IL (ms)} &\text {Memory  (MB)}&\text {E (J)}\\
\hline \text{RN-18} & \textbf{48.2}&\textbf{23}&\textbf{0.26}\\
\text{GN} & 64.1&26&0.37\\
\text{MN v2} & 51.2&38&0.33\\
\hline
\end{array}$
    \end{tabular}
    \vspace{-0.3cm}
    \caption{The inference latency (IL), memory allocated (Memory), and the energy consumed per frame (E) by each model on a smartphone.}
    \vspace{-0.25cm}
    \label{tab:app-all-time}
\end{table}

\vspace{-0.5cm}
\section{Conclusion}
\vspace{-0.3cm}
This paper explores the feasibility of using privacy-preserving, decentralized models for assessing depression and its severity using speech. Toward this goal, we use federated learning to enable on-device training. We used an existing dataset (DAIC-WOZ) to establish a performance benchmark for decentralized and privacy-preserving approaches. We show that the privacy-preserving FL models perform robustly with only 4-6\% accuracy lost compared to a centralized approach. More importantly, these models achieve better accuracy than the best-performing models in prior work using DAIC-WOZ (e.g., 87\% (FL) compared to 74.64\% \cite{srimadhur2020end}). These findings establish the feasibility of using privacy-preserving FL models for depression assessment. We also explored the feasibility of deploying on devices with low computational resources. The FL models show small inference latency and a low memory footprint while being energy-efficient. As such, these models can be deployed to mobile devices to support continuous and real-time assessment of depression at scale.
\vspace{-0.2cm}

\bibliographystyle{IEEEbib}
\vspace{-0.1cm}
\bibliography{refs}

\end{document}